# Information theory explanation of the fluctuation theorem, maximum entropy production and self-organized criticality in non-equilibrium stationary states

**Roderick Dewar**

Unité de Bioclimatologie, INRA Centre de Bordeaux, BP 81, 33883 Villenave d'Ornon CEDEX, France

Email : dewar@bordeaux.inra.fr

**Abstract**

Jaynes' information theory formalism of statistical mechanics is applied to the stationary states of open, non-equilibrium systems. First it is shown that the probability distribution $p_\Gamma$ of the underlying microscopic phase space trajectories $\Gamma$ over a time interval of length $\tau$ satisfies $p_\Gamma \propto \exp(\tau \sigma_\Gamma / 2k_B)$ where $\sigma_\Gamma$ is the time-averaged rate of entropy production of $\Gamma$. Three consequences of this result are then derived : (1) the Fluctuation Theorem, which describes the exponentially declining probability of deviations from the 2[nd] law of thermodynamics as $\tau \to \infty$ ; (2) the selection principle of maximum entropy production for non-equilibrium stationary states, empirical support for which has been found in studies of phenomena as diverse as the Earth's climate and crystal growth morphology ; and (3) the emergence of self-organized criticality for flux-driven systems in the slowly-driven limit. The explanation of these results on general information theoretic grounds underlines their relevance to a broad class of stationary, non-equilibrium systems. In turn, the accumulating empirical evidence for these results lends support to Jaynes' formalism as a common predictive framework for equilibrium and non-equilibrium statistical mechanics.

PACS numbers : 05.70 Ln, 65.40 Gr, 89.70+c

## 1. Introduction

More than 100 years since the development of equilibrium statistical mechanics (ESM) by Boltzmann and Gibbs, the general consensus appears to be that we still do not have an equivalent formalism for predicting the behaviour of non-equilibrium systems. And yet almost 50 years ago, following the development of information theory by Shannon [1], Jaynes [2-3] proposed that Gibbs' approach could provide a common predictive framework for equilibrium and non-equilibrium behaviour.

Gibbs' formalism of ESM consists of maximising the quantity $-\sum_i p_i \ln p_i$ with respect to the microstate probabilities $p_i$, subject to the relevant external constraints on the system – thus yielding the microcanonical, canonical and grand-canonical distributions for isolated, closed and open systems, respectively. In the light of Shannon's interpretation of $-\sum_i p_i \ln p_i$ as a measure of missing information [1], Jaynes [2-3] saw Gibbs' formalism of ESM as but one example of a general form of statistical inference ('Maxent') which could be extended to non-equilibrium systems, as well as to other problems requiring prediction from insufficient data (*e.g.* image reconstruction, spectral analysis, inverse problems). By 1962 Jaynes had shown how the Maxent method led to the known results of linear transport theory (Onsager, Kubo and others) for near-equilibrium systems [4]. However, Jaynes' approach has yet to be generally accepted as a formalism for non-equilibrium statistical mechanics (NESM), due perhaps to the paucity of published results by the 'Maxent school' [5], particularly with regard to new testable predictions of behaviour far from equilibrium.

Now, however, from a variety of empirical and theoretical studies within the last 25 years, three generic features of non-equilibrium behaviour have emerged which offer further opportunities to test Jaynes' formalism of NESM. Firstly, since the 1970s, Paltridge [6-9] and subsequent authors [10-15] have suggested that the observed mean-annual configuration of horizontal energy fluxes in the Earth's atmosphere and oceans corresponds closely to a state of maximum global entropy production. A selection principle of maximum entropy production (MaxEP) also appears to govern the observed vertical heat fluxes in the atmosphere [16, 17]. Wider empirical support for MaxEP has been found in studies of such diverse phenomena as the Earth's thermohaline circulation [18], the zonal climates of Mars and Titan [19], crystal growth morphology [20-21] and bacterial metabolism [22]. A principle of MaxEP was also postulated by Sawada [23]. Some qualitative arguments for MaxEP have been put forward [8-9, 24-25]. In addition, Ziegler's theorem of maximum dissipation [26] may offer a phenomenological approach to understanding MaxEP [15]. However, what is currently lacking is a statistical mechanical framework from which MaxEP emerges as an explicit quantitative prediction.

Secondly, and more recently, an apparently independent property of entropy production, the Fluctuation Theorem (FT), has been established for a variety of non-equilibrium systems [27-32]. The FT relates the probability $p(\sigma_\tau)$ of observing a phase-space trajectory with a whole-system entropy production rate of $\sigma_\tau$ over time interval $\tau$, to that of observing a trajectory with entropy production rate of $-\sigma_\tau$: specifically, $p(\sigma_\tau)/p(-\sigma_\tau) = \exp(\tau\sigma_\tau/k_B)$. This result describes how the probability of violations of the 2nd law of thermodynamics becomes exponentially small as $\tau$ or the system size increases. The FT has been demonstrated mainly through theoretical studies and computer simulations of microscopic dynamics [27-31]. The first experimental demonstration of the FT was given recently by Wang *et al.* [32]. The validity of the FT for a wide class of microscopic systems suggests that it has some generic explanation. Maes [33] has shown that the FT can be understood on the basis of the hypothesis that the underlying microscopic space-time trajectories have a Gibbs-type distribution. While the Gibbs property has been demonstrated in specific cases, its justification on general grounds remains to be established.

A third line of research has led to the emergence of the paradigm of self-organized criticality (SOC) [34-35]. SOC describes the observation that non-equilibrium systems as diverse as earthquakes, sandpiles and biological evolution tend to organize themselves into states whose dynamics vary abruptly, and which are characterised by power-law behaviour reminiscent of equilibrium systems at their critical points. The critical behaviour of equilibrium systems is well understood, and can be predicted from a renormalisation group analysis of Gibbs' grand-canonical distribution. The reason we do not have an equivalent understanding of SOC is, essentially, that we do not yet know the non-equilibrium analogue of Gibbs' distribution [35].

Jaynes [36], commenting in 1980 on Prigogine's principle of *minimum* entropy production [37], conjectured that perhaps Gibbs' method might be closer in spirit to a principle of *maximum* entropy production (although it is not clear whether Jaynes was aware of Paltridge's work [6-8] up to this time). The purpose of this paper is to show that all three behaviours (MaxEP, FT, SOC) are explicit predictions of Jaynes' formalism of NESM. Specifically, we consider the macroscopic stationary states of open, non-equilibrium systems, and use Jaynes' formalism to construct the probability distribution $p_\Gamma$ for the underlying microscopic phase-space paths $\Gamma$. Thus $p_\Gamma$ is the analogue of Gibbs' grand-canonical distribution (to which $p_\Gamma$ reduces in the equilibrium limit, as shown below). MaxEP, FT and SOC are derived as corollaries of the result for $p_\Gamma$.

The plan of the paper is as follows. Section 2 gives an overview of Jaynes' formalism of NESM, its underlying rationale, and a summary of the main steps involved in the subsequent derivation of MaxEP, FT and SOC. This section also introduces some notation. Section 3 describes in detail the derivation of the key result, an expression for the microscopic path distribution $p_\Gamma$. In section 4 the FT, the principle of MaxEP, and SOC are then derived as corollaries of this result. Some concluding remarks are made in section 5.

## 2. Overview of the Jaynes formalism of NESM

### 2.1 What is the Jaynes formalism ?

The Jaynes formalism of NESM, as applied here, simply consists of maximising the path information entropy $S_I = -\sum_\Gamma p_\Gamma \ln p_\Gamma$ with respect to $p_\Gamma$, subject to the imposed constraints.

All macroscopic quantities may then be calculated as expectation values over the distribution $p_\Gamma$. The imposed constraints include external forcings such as solar radiation input at the top of the atmosphere in the case of a planetary climate system, the input of grains in the case of a sandpile, or the imposed temperature gradient in the case of a Rayleigh-Bénard convection cell. The imposed constraints may also include microscopic conservation laws which determine the spectrum of possible paths $\Gamma$. In the case of stationary macroscopic states, a stationary steady-state constraint will also apply.

### 2.2 Underlying rationale

According to a limit theorem first noted by Shannon [1], the information entropy $S_I = -\sum_{i=1}^{N} p_i \ln p_i$ is the logarithm of the number of outcomes $i$ with non-negligible probability $p_i$ (giving a generalisation of Boltzmann's relation $S = k_B \ln W$). This is easily seen in the extreme cases where $p_i$ is uniform ($S_I = \ln N$) and sharply peaked on a single outcome ($S_I = \ln 1 = 0$). In the case of NESM, therefore, $S_I = -\sum_\Gamma p_\Gamma \ln p_\Gamma$ is the logarithm of the number of microscopic phase-space paths $\Gamma$ having non-negligible probability $p_\Gamma$. Jaynes' procedure amounts, then, to finding the most probable macroscopic history, *i.e.* the one that can be realised by the greatest number of microscopic paths (the broadest $p_\Gamma$) compatible with the imposed constraints. The rationale here is analogous to Boltzmann's microstate counting, now in terms of microscopic paths. Paths rather than states are the central objects of interest in non-equilibrium systems, because of the presence of non-zero macroscopic fluxes whose statistical description requires us to consider the underlying microscopic behaviour over time.

An equivalent, more information-based way of viewing the rationale for Jaynes' formalism is to recognise that we are concerned with the prediction of *reproducible* macroscopic behaviour [3]. Whereas the macroscopic behaviour is experimentally reproducible under the applied constraints, the microscopic path $\Gamma$ is not (because we can not set up the system with atomic precision). That the macroscopic behaviour *is* reproducible under given constraints implies that it is characteristic of each of the vast number of microscopic paths compatible with those constraints [3]. It follows that the vast majority of the microscopic details of $\Gamma$ are irrelevant to the prediction of macroscopic quantities. Jaynes' procedure of maximising the information entropy $S_I$, subject *only* to the imposed constraints, effectively discards *all* of the irrelevant microscopic information from $p_\Gamma$. It ensures that the only information encoded in $p_\Gamma$ are the external constraints, the steady-state constraint, and any other *a priori* information such as local conservation laws – we simply do not need any other information if these constraints are sufficient to reproduce the macroscopic behaviour of the system. When applications of the Jaynes procedure fail, the situation is informative because it signals the presence of new constraints that had not been taken into account.

*2.3 Main steps involved*

First we introduce some notation. We will consider a general open system (volume $V$, boundary $\Omega$) exchanging both energy and matter with its surroundings. The system may consist of several constituents ($i = 1\ldots. m$) undergoing mutual transformations. Let $u(\boldsymbol{x},t)$ denote the internal energy density and $\rho_i(\boldsymbol{x},t)$ the mass density of constituent $i$ at position $\boldsymbol{x}$ and time $t$, and let $\boldsymbol{f}_u(\boldsymbol{x},t)$ and $\boldsymbol{f}_i(\boldsymbol{x},t)$ denote the corresponding internal energy and mass flux densities. It is convenient to introduce the macroscopic state vectors $\boldsymbol{d} = (u, \{\rho_i\})$ and $\boldsymbol{F} = (f_u, \{f_i\})$. The normal components of $\boldsymbol{F}$ on the boundary $\Omega$ are denoted by $\boldsymbol{F}^n$. For any quantity $X$, let $\overline{X} = \dfrac{1}{\tau} \int_0^\tau X(t)\mathrm{d}t$ denote its time average over interval $\tau$, and let $\langle X \rangle = \sum_\Gamma p_\Gamma X_\Gamma$ denote its expectation value over the probability distribution $p_\Gamma$ of the underlying microscopic paths, where $X_\Gamma$ is the value of $X$ for path $\Gamma$.

We will use Jaynes' formalism to find the most probable macroscopic stationary state adopted by the system under the imposed constraints. The overall procedure, then, will be to maximise the path information entropy

$$S_I = -\sum_\Gamma p_\Gamma \ln p_\Gamma \qquad (1)$$

with respect to $p_\Gamma$, subject to the external constraints (*e.g.* surface flux inputs, externally imposed gradients), the stationarity constraint (steady-state energy and mass balance), and any other *a priori* information (*e.g.* local energy and mass conservation). We will carry out this procedure in two steps.

***Step 1*** : We will consider a finite time interval from $t = 0$ to $t = \tau$. We will first maximise $S_I$ with respect to $p_\Gamma$, subject to fixed initial configurations of internal energy and mass density $\langle \boldsymbol{d}(\boldsymbol{x},0) \rangle$ within volume $V$ at time $t = 0$, and fixed time-averaged configurations of internal energy and mass flux densities $\left\langle \overline{\boldsymbol{F}^n(\boldsymbol{x})} \right\rangle$ on the boundary $\Omega$ over the subsequent time interval $\tau$. Note that if the system is in a stationary state, then $\langle \boldsymbol{d}(\boldsymbol{x},0) \rangle$ and $\left\langle \overline{\boldsymbol{F}^n(\boldsymbol{x})} \right\rangle$ are sufficient to describe the macroscopic state of the system and its interaction with the surroundings throughout the interval $\tau$. We will also include here the constraints of local energy and mass conservation. This step introduces Lagrange multipliers $\lambda(\boldsymbol{x})$ conjugate to $\langle \boldsymbol{d}(\boldsymbol{x},0)\rangle$ and $\left\langle \overline{\boldsymbol{F}^{\,n}}(\boldsymbol{x})\right\rangle$, and leads to a path distribution $p_\Gamma(\boldsymbol{\lambda})$ and corresponding maximised information entropy $S_{I,\max}(\boldsymbol{\lambda})$ that are functionals of $\boldsymbol{\lambda}(\boldsymbol{x})$. It is at this point that we will see the entropy production of path $\Gamma$ emerge as a key determinant of $p_\Gamma(\boldsymbol{\lambda})$. The Fluctuation Theorem will follow as a direct consequence of the result for $p_\Gamma(\boldsymbol{\lambda})$.

***Step 2*** : We will then maximise $S_{I,\max}(\boldsymbol{\lambda})$ with respect to $\boldsymbol{\lambda}(\boldsymbol{x})$ subject to the remaining constraints, *e.g.* external constraints and stationarity constraint. The logic here is that the values of the unknown averages $\langle \boldsymbol{d}(\boldsymbol{x},0)\rangle$ and $\left\langle \overline{\boldsymbol{F}^{\,n}}(\boldsymbol{x})\right\rangle$ that were artificially fixed during Step 1 are, for a macroscopically reproducible state, ultimately determined only by the remaining constraints. In this step, therefore, we are discarding the residual irrelevant information in $p_\Gamma(\boldsymbol{\lambda})$ due to having fixed $\langle \boldsymbol{d}(\boldsymbol{x},0)\rangle$ and $\left\langle \overline{\boldsymbol{F}^{\,n}}(\boldsymbol{x})\right\rangle$. The solution for $\boldsymbol{\lambda}$ will be denoted $\boldsymbol{\lambda}^*$. The resulting distribution $p_\Gamma(\boldsymbol{\lambda}^*)$ completes the Jaynes procedure. MaxEP will be shown to be equivalent to this second step, and SOC will emerge as a special case of MaxEP applied to flux-driven systems in the slowly-driven limit.

## 3. Derivation of path distribution $p_\Gamma(\boldsymbol{\lambda})$

This section describes the derivation of $p_\Gamma(\boldsymbol{\lambda})$, completing Step 1 above. Together with normalisation of $p_\Gamma$, the configurations $\langle \boldsymbol{d(x,0)} \rangle$ and $\langle \overline{\boldsymbol{F^n}(x)} \rangle$ yield the following constraints on $p_\Gamma$:

$$\sum_\Gamma p_\Gamma = 1 \qquad\qquad\qquad\qquad (2)$$

$$\sum_\Gamma p_\Gamma \boldsymbol{d(x,0)}_\Gamma = \langle \boldsymbol{d(x,0)} \rangle \qquad\qquad (x \in V) \qquad\qquad (3)$$

$$\sum_\Gamma p_\Gamma \overline{\boldsymbol{F^n}(x)}_\Gamma = \langle \overline{\boldsymbol{F^n}(x)} \rangle \qquad\qquad (x \in \Omega) . \qquad\qquad (4)$$

Now maximising $S_I$ [equation (1)] with respect to $p_\Gamma$, subject to equations (2)-(4), yields the distribution

$$p_\Gamma = \frac{1}{Z} \exp(A_\Gamma) \qquad\qquad\qquad\qquad (5)$$

in which $Z = \sum_\Gamma \exp(A_\Gamma)$ is a normalisation factor (partition function) and $A_\Gamma$ is the path action given by

$$A_\Gamma = \int_V \boldsymbol{\lambda(x)} \cdot \boldsymbol{d(x,0)}_\Gamma + \int_\Omega \boldsymbol{\eta(x)} \cdot \overline{\boldsymbol{F^n}(x)}_\Gamma \qquad\qquad (6)$$

where $\boldsymbol{\lambda(x)}$ and $\boldsymbol{\eta(x)}$ are Lagrange multipliers determined from equations (3) and (4) by the functional relationships

$$\frac{\delta \ln Z}{\delta \boldsymbol{\lambda(x)}} = \langle \boldsymbol{d(x,0)} \rangle \qquad\qquad (x \in V) \qquad\qquad (7)$$

$$\frac{\delta \ln Z}{\delta \eta(x)} = \langle \overline{\boldsymbol{F^n}(x)} \rangle \qquad\qquad (x \in \Omega) . \qquad\qquad (8)$$

The multipliers $\boldsymbol{\lambda(x)}$ and $\boldsymbol{\eta(x)}$ are not independent on the boundary $\Omega$, being linked there through the constraints of local internal energy and mass balance :

$$\frac{\partial \boldsymbol{d}(\boldsymbol{x},t)_\Gamma}{\partial t} = -\nabla \cdot \boldsymbol{F}_\Gamma + \boldsymbol{Q}_\Gamma \qquad (9)$$

in which $\boldsymbol{Q}$ is the source term for $\boldsymbol{d}$. For a multi-component fluid with chemical reactions, $\boldsymbol{Q}$ has components ($\sum_{mn} \phi_{mn} \frac{\partial v_m}{\partial x_n}$, { $\sum_r v_{ir} j_r$ }), where $\phi_{mn}$ is the stress tensor, $\frac{\partial v_m}{\partial x_n}$ is the rate-of-strain tensor ( $v$ = barycentric velocity), with $v_{ir}$ and $j_r$ the stoichiometric coefficients and rate of reaction $r$ respectively. In order to introduce these constraints into equation (6), we write

$$\boldsymbol{d}(\boldsymbol{x},0)_\Gamma = \frac{1}{2}\big[\boldsymbol{d}(\boldsymbol{x},0)_\Gamma + \boldsymbol{d}(\boldsymbol{x},\tau)_\Gamma\big] - \frac{\tau}{2}\overline{\frac{\partial \boldsymbol{d}(\boldsymbol{x},t)_\Gamma}{\partial t}} \qquad (10)$$

and substitute $\frac{\partial \boldsymbol{d}(\boldsymbol{x},t)_\Gamma}{\partial t}$ from equation (9). After an integration by parts on $\boldsymbol{x}$, equation (6) then becomes (omitting $\boldsymbol{x}$ for brevity)

$$A_\Gamma = \frac{1}{2}\int_V \boldsymbol{\lambda}\cdot\big(\boldsymbol{d}_{\Gamma(0)} + \boldsymbol{d}_{\Gamma(\tau)}\big) - \frac{\tau}{2}\int_V \big(\overline{\boldsymbol{F}_\Gamma}\cdot\nabla\boldsymbol{\lambda} + \boldsymbol{\lambda}\cdot\overline{\boldsymbol{Q}_\Gamma}\big) \; + \int_\Omega \left(\frac{\boldsymbol{\lambda}\tau}{2}+\boldsymbol{\eta}\right)\cdot\overline{\boldsymbol{F}_\Gamma^n} \qquad (11)$$

in which $\boldsymbol{d}_{\Gamma(0)} = \boldsymbol{d}(\boldsymbol{x},0)_\Gamma$ and $\boldsymbol{d}_{\Gamma(\tau)} = \boldsymbol{d}(\boldsymbol{x},\tau)_\Gamma$. Because information about the fluxes throughout the system is now also contained in the term $\int_V \overline{\boldsymbol{F}_\Gamma}\cdot\nabla\boldsymbol{\lambda}$, the information encoded in the surface integral is redundant and the corresponding Lagrange multiplier ½$\boldsymbol{\lambda}\tau + \boldsymbol{\eta}$ can be set to zero. Thus

$$A_\Gamma(\boldsymbol{\lambda}) = \frac{1}{2}\int_V \boldsymbol{\lambda}\cdot\big(\boldsymbol{d}_{\Gamma(0)} + \boldsymbol{d}_{\Gamma(\tau)}\big) - \frac{\tau}{2}\int_V \big(\overline{\boldsymbol{F}_\Gamma}\cdot\nabla\boldsymbol{\lambda} + \boldsymbol{\lambda}\cdot\overline{\boldsymbol{Q}_\Gamma}\big) \qquad (12)$$

where we indicate the dependence of the action on $\boldsymbol{\lambda}$ explicitly, $p_\Gamma(\boldsymbol{\lambda})$ then being given by equation (5).

Finally, a more familiar expression for $A_\Gamma(\boldsymbol{\lambda})$ is obtained by defining the local temperature and chemical potentials ($T$ and $\mu_i$) through

$$\lambda = \frac{1}{k_B T} \ \left(-1, \{\mu_i\}\right) \tag{13}$$

where $k_B$ is Boltzmann's constant. Equation (12) becomes

$$A_\Gamma \left(T, \{\mu_i\}\right) = -\frac{1}{2} \int_V \frac{H_{\Gamma(0)} + H_{\Gamma(\tau)}}{k_B T} \ + \ \frac{\tau \sigma_\Gamma}{2 k_B} \tag{14}$$

where $H = u - \sum_i \mu_i \rho_i$ is the non-equilibrium generalisation of the grand-canonical Hamiltonian, $H_{\Gamma(0)}$ and $H_{\Gamma(\tau)}$ are the values of $H$ at the end points of trajectory $\Gamma$ ($t = 0$ and $\tau$), and $\sigma_\Gamma$ is the time-averaged rate of entropy production of $\Gamma$ defined by

$$\sigma_\Gamma = \int_V \left( \overline{f_{u(\Gamma)}} \cdot \nabla\left(\frac{1}{T}\right) - \sum_i \overline{f_{i(\Gamma)}} \cdot \nabla\left(\frac{\mu_i}{T}\right) + \frac{1}{T} \sum_{mn} \overline{\phi_{mn(\Gamma)} \frac{\partial v_{m(\Gamma)}}{\partial x_n}} - \frac{1}{T} \sum_{ir} \mu_i \nu_{ir} \overline{j_{r(\Gamma)}} \right) \tag{15}$$

in the case of a multi-component fluid. Equation (12) provides a more general definition of the dimensionless entropy production ($\Sigma_\Gamma$) in terms of fluxes $\boldsymbol{F}$ and sources $\boldsymbol{Q}$ through

$$\Sigma_\Gamma = -\ \tau \int_V \left( \overline{F_\Gamma} \cdot \nabla\lambda + \lambda \cdot \overline{Q_\Gamma} \right) \quad .$$

Equations (5) and (12) [or (14)] constitute the central result of Step 1, which may be summarised thus : for a non-equilibrium system characterised by given configurations of $\langle \boldsymbol{d}(\boldsymbol{x},0) \rangle$ in $V$ and $\left\langle \overline{\boldsymbol{F}^n}(\boldsymbol{x}) \right\rangle$ on $\Omega$, and obeying local energy and mass conservation, the probability $p_\Gamma(\lambda)$ of a microscopic path $\Gamma$ depends on the grand-canonical Hamiltonian $H$ at its end points and on the entropy production $\tau \sigma_\Gamma$ of the intermediate pathway.

Before examining the implications of this result in the following section, let us consider the physical basis of the above definitions of temperature, chemical potential and entropy production for non-equilibrium systems. We have seen how the constrained variational problem of maximising the path information entropy $S_I$ leads to a natural definition of $T$ and $\mu_i$ in terms of the Lagrange multipliers conjugate to the internal energy and mass densities ($<u>$ and $<\rho_i>$). Because $u$ and $\rho_i$ are physical observables whose definitions do not depend on the existence of equilibrium, $T$ and $\mu_i$ are likewise physically well-defined out of equilibrium. The actual definition of $T$ and $\mu_i$ through equation (13) ensures the correct correspondence with the equilibrium limit, in which the path distribution reduces to Gibbs' grand-canonical distribution.

Also, the definition of the entropy production rate given by equation (15) is consistent with the definition of the non-equilibrium entropy density $s$ (not to be confused with the information entropy $S_I$) given by the increment relationship

$$T\delta s = \delta u - \sum_i \mu_i \delta \rho_i \quad . \tag{16}$$

This may be verified by dividing equation (16) through by d$t$ and invoking internal energy and mass balance as in equation (9), to find the entropy balance equation

$$\int_V \overline{\frac{\partial s_\Gamma}{\partial t}} = \sigma_\Gamma - \int_\Omega \frac{1}{T}\left( \overline{f_{u(\Gamma)}^n} - \sum_i \mu_i \overline{f_{i(\Gamma)}^n} \right) \quad , \tag{17}$$

in which $\sigma_\Gamma$ is given by equation (15). The right-hand side of equation (17) may then be interpreted as the difference between the time-averaged rates of entropy production within volume $V$ and entropy export across the boundary $\Omega$, for trajectory $\Gamma$ over interval $\tau$. To summarise, the quantities $T$, $\mu_i$, $s$ and $\sigma_\Gamma$ are physically well-defined however far the system is from equilibrium, because they are entirely defined through $u$ and $\rho_i$ and their respective conservation laws. In the information theoretic approach, therefore, there is no need to invoke a local thermodynamic equilibrium hypothesis to define these quantities [cf. 38].

## 4. Three implications

### 4.1 Second law and fluctuation theorem

Under replacement of the microscopic path $\Gamma$ by its time-reversal $\Gamma_R$, the end-point contribution to the path action $A_\Gamma$ is symmetric whereas the contribution from the pathway entropy production is asymmetric ($\sigma_{\Gamma_R} = -\sigma_\Gamma$). Equations (5) and (14) therefore imply that the ratio of the probabilities of $\Gamma$ and $\Gamma_R$ is given by

$$p_\Gamma / p_{\Gamma_R} = \exp(\tau \sigma_\Gamma / k_B) \qquad (18)$$

For an arbitrary path distribution $p_\Gamma$ the function $Y(\theta) = \left\langle \left( p_\Gamma / p_{\Gamma_R} \right)^{-\theta} \right\rangle$ satisfies

$$Y(\theta) = \sum_\Gamma p_\Gamma^{1-\theta} p_{\Gamma_R}^{\theta} = \sum_\Gamma p_{\Gamma_R}^{1-\theta} p_\Gamma^{\theta} = Y(1-\theta) \qquad (19)$$

through exchange of dummy indices $\Gamma$ and $\Gamma_R$. In the particular case of equation (18) we can use this result with $\theta = 1$ to find $\left\langle \exp(-\tau \sigma / k_B) \right\rangle = 1$ which, by Jensen's inequality $\exp(\langle x \rangle) \leq \langle \exp x \rangle$, implies that

$$\langle \sigma \rangle \geq 0 \qquad (20)$$

*i.e.* the 2$^{\text{nd}}$ law of thermodynamics is satisfied on average. It also follows from equation (18) that the probability distribution of the entropy production $p(\sigma) = \sum_\Gamma \delta(\sigma - \sigma_\Gamma) p_\Gamma$ satisfies the

Fluctuation Theorem

$$p(\sigma) = \exp(\tau \sigma / k_B) p(-\sigma) \quad . \qquad (21)$$

which describes how the probability of violations of the 2$^{\text{nd}}$ law becomes exponentially small as $\tau$ or the system size increases. These results are a direct consequence of the Gibbs-type form for $p_\Gamma(\lambda)$ [cf. 33], whose origin is explained here on general grounds as the result of Step 1 of the Jaynes procedure for maximising the path information entropy.

*4.2 Maximum entropy production*

An expression for the maximised path information entropy $S_{I,\max}(\boldsymbol{\lambda})$ after Step 1 is obtained by substituting equation (5) into equation (1), giving

$$S_{I,\max}(\boldsymbol{\lambda}) = -\sum_{\Gamma} p_{\Gamma}(\boldsymbol{\lambda}) \ln p_{\Gamma}(\boldsymbol{\lambda}) = \ln Z(\boldsymbol{\lambda}) - \langle A(\boldsymbol{\lambda}) \rangle \tag{22}$$

Recall that the path action $A_{\Gamma}(\boldsymbol{\lambda}) = A_{\Gamma}^{\text{rev}}(\boldsymbol{\lambda}) + A_{\Gamma}^{\text{irr}}(\boldsymbol{\lambda})$ is the sum of a reversible part ($A_{\Gamma}^{\text{rev}}(\boldsymbol{\lambda})$, the end-point contribution) which is symmetric under path reversal ($\Gamma \rightarrow \Gamma_{R}$), and an irreversible part ($A_{\Gamma}^{\text{irr}}(\boldsymbol{\lambda}) = \tau \sigma_{\Gamma} / 2 k_{B}$, the contribution from the path entropy production) which is asymmetric under path reversal.

Let us consider the number of paths $\Gamma$ which contribute to the mean behaviour, and which therefore have reversible action $\langle A^{\text{rev}}(\boldsymbol{\lambda}) \rangle$ and irreversible action $\langle A^{\text{irr}}(\boldsymbol{\lambda}) \rangle$. Note that for every such path $\Gamma$, there exists another path (the reversed path $\Gamma_{R}$ with the same reversible action $\langle A^{\text{rev}}(\boldsymbol{\lambda}) \rangle$ but an irreversible action of $-\langle A^{\text{irr}}(\boldsymbol{\lambda}) \rangle$. Therefore the number of paths contributing to the mean behaviour is $W(\langle A^{\text{irr}}(\boldsymbol{\lambda}) \rangle)$, the number of paths with irreversible action equal to $\langle A^{\text{irr}}(\boldsymbol{\lambda}) \rangle$. Ignoring fluctuations about the mean behaviour, we then have the approximation

$$Z(\boldsymbol{\lambda}) = \sum_{\Gamma} \exp(A_{\Gamma}(\boldsymbol{\lambda})) \approx W(\langle A^{\text{irr}}(\boldsymbol{\lambda}) \rangle) \exp(\langle A(\boldsymbol{\lambda}) \rangle) \tag{23}$$

Equations (22) and (23) then yield the result

$$S_{I,\max}(\boldsymbol{\lambda}) \approx \ln W(\langle A^{\text{irr}}(\boldsymbol{\lambda}) \rangle) \tag{24}$$

Assuming that $W(A^{\text{irr}})$ is an increasing function of $A^{\text{irr}}$, then Step 2 of the Jaynes procedure, in which we maximise $S_{I,\max}(\boldsymbol{\lambda})$ with respect to $\boldsymbol{\lambda}(x)$ subject to the remaining constraints (external constraints, stationarity constraint), is equivalent to maximising the mean entropy production rate $\langle \sigma(\boldsymbol{\lambda}) \rangle$ with respect to $\boldsymbol{\lambda}(x)$, subject to those same constraints. That is, within the mean-field approximation that neglects fluctuations in the action, Step 2 of the Jaynes procedure is equivalent to the principle of MaxEP.

As equation (15) illustrates, the entropy production rate can be interpreted as a product of fluxes and their conjugate forces. In these terms, a maximum in entropy production exists because, under the steady-state constraints of energy and mass balance, the fluxes act to dissipate the conjugate driving forces, leading to a trade-off between fluxes and forces. A toy example, in the spirit of Paltridge [9], serves to illustrate this trade-off. Consider a two-box climate system (box 1 = 'equatorial zone' at temperature $T_1$, box 2 = 'polar zone' at temperature $T_2$), with short-wave radiation input $f_{SW}$ into box 1 only, long-wave radiation outputs from box $i$ proportional to $T_i^4$, and meridional heat flux $h$ from box 1 to box 2. Steady-state energy balance then implies $T_1 \propto (f_{SW}-h)^{1/4}$ and $T_2 \propto h^{1/4}$. As a result the entropy production rate $\sigma = h(1/T_2 - 1/T_1)$ has a maximum with respect to $h$ at $h_{opt} \approx 0.199 f_{SW}$, between the values $h = 0$ (zero flux) and $h = 0.5 f_{SW}$ (zero driving force) at which $\sigma$ vanishes. The principle of MaxEP predicts that $h_{opt}$ is the actual meridional heat flux which is physically selected.

One objection to Paltridge's work has been that his formulation of MaxEP for the Earth's climate only includes contributions to global entropy production from horizontal heat transport in the atmosphere and oceans, and ignores the radiative entropy change associated with the transfer of solar radiation from solar to terrestrial temperatures [39]. In this regard it should be noted that Jaynes' procedure tells us which quantity is maximised, namely the rate of entropy production $\sigma_\Gamma$ defined through equations (12) and (14), as derived from the local internal energy and mass balance of the system (equation (9)). If we apply Jaynes' procedure to the matter component of Earth's climate (effectively treating the radiation component as an imposed constraint), then $\sigma_\Gamma$ is the material entropy production given by equation (15), which includes contributions from both horizontal and vertical heat transport, but does not involve the radiative entropy production. The radiative entropy production is then irrelevant when applying MaxEP to the climate system (as in Paltridge's formulation), in agreement with the entropy balance analyses of Goody and Abdou [40, 41]. This conclusion might be examined more fully through an application of Jaynes' procedure to the combined system of matter and radiation, along the lines of [42].

For climate systems there is an additional contribution to $\sigma_T$ from radiative heating at planetary temperatures, deriving from the internal energy component of $\boldsymbol{Q}$ in equations (9) and (12). This contribution is not taken into account in the purely material entropy production of equation (15), and is in fact ignored in applications of MaxEP to the climate [16, 17]. Again Jaynes' procedure provides the rationale for ignoring this contribution – radiative heating is reversible and does not contribute to the number of paths $W\left(\left\langle A^{\mathrm{irr}}(\boldsymbol{\lambda})\right\rangle\right)$ in equation (24). To summarise, in applications of MaxEP to climate systems it is the irreversible, material entropy production that is maximised.

### 4.3 Self-organized criticality

We have seen how Step 2 of the Jaynes procedure gives (within a mean-field approximation) a path distribution $p_T(\boldsymbol{\lambda})$ for which the corresponding mean entropy production rate is maximised with respect to $\boldsymbol{\lambda}$, subject to the external constraints on the system (and the stationarity constraint). Denoting by $\boldsymbol{\lambda}^*$ the solution for $\boldsymbol{\lambda}$ that satisfies the MaxEP principle, all macroscopic quantities reproduced under the external constraints can be calculated as expectation values over $p_T(\boldsymbol{\lambda}^*)$. The MaxEP state ($\boldsymbol{\lambda} = \boldsymbol{\lambda}^*$) is, in other words, the reproducible state into which the system self-organises under the given external constraints.

Let us now consider a flux-driven system in a MaxEP state, for example a sandpile driven by a fixed external grain influx of magnitude $F_{\mathrm{ext}}$. Let $F$ be the magnitude of the total outflux of grains. The stationarity constraint implies that $\langle F \rangle = F_{\text{ext}}$. For the sandpile, $\boldsymbol{\lambda}(\boldsymbol{x})$ would be the chemical potential conjugate to the mean number of grains at $\boldsymbol{x}$, and thus would be a function of the mean local height $h(\boldsymbol{x})$, with $\boldsymbol{\lambda}^*(\boldsymbol{x})$ corresponding to the mean height profile into which the sandpile self-organises for given $F_{\text{ext}}$.

In principle we could use $p_\Gamma(\boldsymbol{\lambda}^*)$ to construct the output flux probability distribution $p(F)$, including its dependence on $F_{\text{ext}}$, through

$$p(F|F_{\text{ext}}) = \sum_\Gamma \delta(F - F_\Gamma) p_\Gamma(\boldsymbol{\lambda}^*) \tag{25}$$

Instead let us use symmetry considerations to construct a Landau-Ginzburg approximation for $p(F)$ in the slow-driving limit $F_{\text{ext}} \to 0$. The objective here is to show in a general way how SOC emerges from MaxEP within the mean-field approximation, rather than to describe a specific system in detail for which a calculation based explicitly on equation (25) might be required.

Specifically, we will suppose that

$$p(F|F_{\text{ext}}) = \frac{1}{Z(F_{\text{ext}})} \exp\{H(F|F_{\text{ext}})\} \tag{26}$$

where $Z(F_{ext}) = \int\limits_0^\infty dF \exp\{H(F|F_{ext})\}$ and, for small $F$,

$$H(F|F_{ext}) = rF^2 + gF^4 \tag{27}$$

where the expansion in even powers reflects the fact that paths contributing to a given magnitude $F$ do so in pairs $(\Gamma, \Gamma_R)$ related by time reversal, each pair having weight $\exp(A_\Gamma(\boldsymbol{\lambda}^*)) + \exp(-A_\Gamma(\boldsymbol{\lambda}^*))$; we ignore the end-point contribution in the limit $\tau \to \infty$. We will take $r > 0$ and $g < 0$. A steepest-descent calculation then yields $\langle F \rangle = \sqrt{-r/2g}$ and from the stationarity constraint $\langle F \rangle = F_{\text{ext}}$ we identify $r = -2gF_{\text{ext}}^2$. Small fluctuations in $F$ around $\langle F \rangle$ are governed by the value of $H'' = d^2H/dF^2$ at $\langle F \rangle$. Specifically, the mean-field approximation to the variance of $F$ is given by

$$\left\langle \left( F - \langle F \rangle \right)^2 \right\rangle \approx \frac{1}{|H''|} = \frac{1}{8|g|F_{ext}^2} \tag{28}$$

which diverges as $F_{ext} \to 0$. Thus the characteristic signal of SOC emerges in the slow-driving limit.

The usual weaknesses inherent in mean-field analyses of SOC [35] also apply here, so that equation (28) is likely to be only qualitatively correct. However, the main point has been to show how Jaynes' formalism allows us to understand SOC as being the macroscopic behaviour which is reproducible (*i.e.* most probable) under the constraint of a slow driving flux. In principle, through calculations based directly on equation (25), Jaynes' procedure provides a general predictive framework for flux-driven systems, which for example could be used to identify further criteria for the emergence of SOC (*e.g.* threshold dynamics, low friction [35]).

In the case of the one-dimensional sandpile model [43] it is intuively clear that SOC corresponds to a MaxEP state. The mean rate of entropy production is proportional to $F_{ext}(h_{x=L} - h_{x=0})$ which, because $F_{ext}$ is fixed, has a maximum when the average slope adopts its largest possible value, *i.e.* the critical slope $dh/dx|_c$. This behaviour contrasts with that of systems driven by fixed gradients. For such systems the MaxEP state corresponds to a maximum in the conjugate flux [23], consistent with the principles of maximum heat and momentum transport proposed by Malkus and Busse for Bénard-type thermal convection and turbulent shear flow, respectively.

## 5. Concluding remarks

Paltridge [6-7, 9] and Grassl [10] demonstrated the practical benefits of the principle of maximum entropy production for the prediction of climate and climate change, in avoiding the need to represent in detail the vast number of internal processes involved. The explanation of this principle – together with that of the Fluctuation Theorem and SOC – on general information theoretic grounds serves to underline the relevance of these results to a broad class of stationary, non-equilibrium systems. Their wider relevance may be examined within an extension of the present formalism to reproducible non-stationary behaviour (*e.g.* cyclic steady states, approach to equilibrium), in which the Lagrange multipliers $\lambda(x)$ would generalise to $\lambda(x,t)$ [4].

In turn, the accumulating empirical evidence for these results over the last 25 years lends further support to Jaynes' formalism as a common predictive framework for equilibrium and non-equilibrium statistical mechanics [2-4]. In each case the problem is to predict macroscopically reproducible behaviour, and the solution is to maximize the information entropy subject to the imposed constraints. The two cases differ only in the nature of the constraints that are imposed.

Finally, it is worth emphasising the general character of Jaynes' procedure, whose predictions may therefore apply to certain dynamical systems lying outside the immediate scope of physics, such as economies and biological populations. Provided the reproducible stationary states of a system can be described by quantities $d(x,t)$ obeying a local conservation law with fluxes $F$ and sources $Q$ (equation (9)), then the Jaynes procedure leads to the general result for $p_\Gamma(\lambda)$ given by equations (5) and (12), in which a generalised entropy production can be defined through the second term in equation (12). The corollaries of this result (FT, MaxEP, SOC) may then be expected as generic features of such systems.

## Acknowledgements

I am indebted to Michael Raupach for introducing me to Garth Paltridge's work, and for his kind encouragement. I thank Chris Jarzynski, Alan McKane, Tim Lenton, Ralph Lorenz, Hisashi Ozawa, Garth Paltridge, Tony Pujol, Michael Roderick and an anonymous referee for valuable discussions and comments on earlier versions of this paper. HO and TP provided clarifications concerning the application of MaxEP to climate systems.